\documentclass[sigplan,screen,10pt,letterpaper]{acmart}
\usepackage{libertine}
\usepackage{booktabs} 

\usepackage[utf8]{inputenc} 
\usepackage{times}
\usepackage[scaled=0.85]{helvet}
\usepackage{graphicx}
\usepackage{multirow}
\usepackage{ifthen}
\usepackage{xspace}
\usepackage{alltt}
\usepackage{latexsym}
\usepackage{url}
\usepackage{amssymb}
\usepackage{amsfonts}
\usepackage{amsmath}
\usepackage{stmaryrd}
\usepackage{enumerate}
\usepackage{caption}
\usepackage{subcaption}
\usepackage{pifont}
\usepackage{xcolor}
\definecolor{greencolor}{RGB}{50,255,50}

\definecolor{redcolor}{RGB}{255,80,80}

\usepackage{xcolor}
\usepackage[normalem]{ulem} 			
\usepackage{ifthen}

\newboolean{showcomments}
\setboolean{showcomments}{true}

\ifthenelse{\boolean{showcomments}}
{
	\newcommand{\nb}[3]{
		{\colorbox{#2}{\bfseries\sffamily\scriptsize\textcolor{white}{#1}}}
		{\textcolor{#2}{\textsf\small$\blacktriangleright$\textit{#3}$\blacktriangleleft$}}}
	 
	\newcommand{\bnote}[2]{\fbox{\color{blue}\bfseries\sffamily\scriptsize#1}
    	{\color{blue}\textsf\small$\blacktriangleright$\textit{#2}$\blacktriangleleft$}}
	\newcommand{\old}[1]{{\color{gray}\sout{#1}}} 
	\newcommand{\del}[1]{\old{#1}} 
	\newcommand{\ins}[1]{{\textcolor{blue}{\uline{#1}}}} 
	\newcommand{\ugh}[1]{{\textcolor{red}{\uwave{#1}}}} 
	\newcommand{\chg}[2]{{\textcolor{red}{\sout{#1}}}{\ra}\textcolor{blue}{\uline{#2}}} 
	 
	\newcommand{\fix}[1]{\bnote{FIX}{#1}}
}{
	\newcommand{\bnote}[2]{}
	\newcommand{\nb}[3]{}
	\newcommand{\old}[1]{}
	\newcommand{\del}[1]{}
	\newcommand{\ins}[1]{}
	\newcommand{\ugh}[1]{}
	\newcommand{\chg}[2]{}
	
	\newcommand{\fix}[1]{}
} 

\newcommand{\hide}[1]{}

\newcommand{\sd}[1]{\nb{Stef}{green}{#1}}

\newcommand{\td}[1]{\nb{Thomas}{orange}{#1}}

\newcommand{\vince}[1]{\nb{Vince}{purple}{#1}}

\graphicspath{{figures/}}

\newcommand{\commented}[1]{}

\newcommand{\eg}{\emph{e.g.,}\xspace}
\newcommand{\ie}{\emph{i.e.,}\xspace}
\newcommand{\etal}{\emph{et al.,}\xspace}
\newcommand{\ct}[1]{{\textsf{#1}}\xspace}

\usepackage{url}            
\makeatletter
\def\url@leostyle{%
  \@ifundefined{selectfont}{\def\UrlFont{\textsf}}{\def\UrlFont{\small\sffamily}}}
\makeatother
\usepackage{color}
\usepackage{listings}
\usepackage{etoolbox}
\definecolor{stComment}{rgb}{0.5,0.5,0.5}
\definecolor{stString}{rgb}{0.58,0,0.82}
\definecolor{stKeywords}{rgb}{0.21,0.55,0.7}
\definecolor{stNumbers}{rgb}{.5,0,0}

\newtoggle{InString}{}
\togglefalse{InString}
\usepackage{color}
\usepackage{textcomp}
\usepackage{listings}
 
\definecolor{source}{gray}{0.85}
\newcommand{\myCommentStyle}[1]{{\sffamily\color{gray!100!white} #1}}
 
\newcommand{\myStringStyle}[1]{{\sffamily\color{violet!100!black} #1}}
 
\newcommand{\mySymbolStyle}[1]{{\sffamily\color{violet!100!black} #1}}
 
\newcommand{\myKeywordStyle}[1]{{\sffamily\color{green!70!black} #1}}
 
\newcommand{\myGlobalStyle}[1]{{\sffamily\color{blue!100!black} #1}}
 
\makeatletter
\newcommand*\idstyle[1]{%
\expandafter\id@style\the\lst@token{#1}\relax%
}
\def\id@style#1#2\relax{%
\ifnum\pdfstrcmp{#1}{\#}=0%
\mySymbolStyle{\the\lst@token}%
\else%
\edef\tempa{\uccode`#1}%
\edef\tempb{`#1}%
\ifnum\tempa=\tempb%
\myGlobalStyle{\the\lst@token}%
\else%
\the\lst@token%
\fi%
\fi%
}
\makeatother

\lstnewenvironment{code}{%
 \lstset{%
 frame=single,
 framerule=0pt,
 mathescape=false
 }%
 \noindent%
 \minipage{\linewidth}%
}{%
 \endminipage%
}%
\lstnewenvironment{codeWithLineNumbers}{%
 \lstset{%
 frame=single,
 framerule=0pt,
 mathescape=false,
 numbers=left
 }%
 \noindent%
 \minipage{\linewidth}%
}{%
 \endminipage%
}%

\newcommand{\sindarin}{Sindarin\xspace}
\newcommand{\Sindarin}{\sindarin}

\usepackage{booktabs}   
\usepackage{subcaption} 

\setcopyright{acmlicensed}
\begin{document}
\copyrightyear{2019}
\acmYear{2019}
\acmConference[DLS '19]{Proceedings of the 15th ACM SIGPLAN International Symposium on Dynamic Languages}{October 20, 2019}{Athens, Greece}
\acmBooktitle{Proceedings of the 15th ACM SIGPLAN International Symposium on Dynamic Languages (DLS '19), October 20, 2019, Athens, Greece}
\acmPrice{15.00}
\acmDOI{10.1145/3359619.3359745}


\title[]{Sindarin: A Versatile Scripting API for the Pharo Debugger}         


\author{Thomas Dupriez}
\affiliation{
  \institution{Univ. Lille, CNRS, Centrale Lille, Inria, UMR 9189 - CRIStAL}            
  \country{France}                    
}
\email{thomas.dupriez@univ-lille.fr}          

\author{Guillermo Polito}
\affiliation{
	\institution{CNRS - UMR 9189 - CRIStAL, Univ. Lille, Centrale Lille, Inria}            
	\country{France}                    
}
\email{guillermo.polito@inria.fr}          

\author{Steven Costiou}
\affiliation{
  \institution{Inria, Univ. Lille, CNRS, Centrale Lille, UMR 9189 - CRIStAL}            
  \country{France}                    
}
\email{steven.costiou@inria.fr}          

\author{Vincent Aranega}
\affiliation{
  \institution{Univ. Lille, CNRS, Centrale Lille, Inria, UMR 9189 - CRIStAL}            
 \country{France}                    
}
\email{vincent.aranega@univ-lille.fr}          

\author{St\'ephane Ducasse}
\affiliation{
  \institution{Inria, Univ. Lille, CNRS, Centrale Lille, UMR 9189 - CRIStAL}            
  \country{France}            
}
\email{stephane.ducasse@inria.fr}          


\begin{abstract}
	Debugging is one of the most important and time consuming activities in software maintenance, yet mainstream debuggers are not well-adapted to several debugging scenarios.
	This has led to the research of new techniques covering specific families of complex bugs.
	Notably, recent research proposes to empower developers with scripting DSLs, plugin-based and moldable debuggers.
	However, these solutions are tailored to specific use-cases, or too costly for one-time-use scenarios.
	


	
	In this paper we argue that exposing a debugging scripting interface in mainstream debuggers helps in solving many challenging debugging scenarios.
	For this purpose, we present \sindarin, a scripting API that eases the expression and automation of different strategies developers pursue during their debugging sessions.
	\sindarin provides a GDB-like API, augmented with AST-bytecode-source code mappings and object-centric capabilities.
	To demonstrate the versatility of \sindarin, we reproduce several advanced breakpoints and non-trivial debugging mechanisms from the literature.
\end{abstract}




\maketitle

\vspace{-0.35cm}
\section{Introduction}
Debugging is an important part of software development.
Literature describes it as a difficult task, on which developers spend sometimes more than 50\% of their development time~\cite{Vess86a,Tell01a,Tass02a,Zell09a,Spin18a}.
Traditional debuggers, often called breakpoint-based or online debuggers, support interactive debug sessions via a graphical user interface.
In such debuggers, the developer makes a hypothesis about the cwause of a bug and manually places breakpoints in the relevant pieces of the code.
When the debugger stops in a breakpoint, the developer explores the subsequent execution by commanding a \emph{step-by-step} execution on it.

Studies and work on debugging acknowledge that mainstream debuggers are not well adapted to several debugging scenarios~\cite{Lieb97a, Berg11h, Pers17a}.
This has led to the appearance of new debugging techniques proposing to augment traditional interactive debuggers with, \eg stateful breakpoints~\cite{Bodd17a}, control-flow aware breakpoints~\cite{Cher07c}, object-centric breakpoints~\cite{Ress12a, Corr01a}, the automatic insertion of breakpoints based on dynamic executions~\cite{Zhan10a}, or declarative statements from the developer~\cite{Ko08a}.
A line of research has also started to study scripting APIs to empower developers to implement debugging scripts adapted to their needs.
These scripting APIs help developers to \eg insert breakpoints based on cross-cutting concerns~\cite{Haih13a} or declaratively explore post-mortem executions~\cite{Phan13a}.
Even further, the moldable debugger framework~\cite{Chis15c} allows developers to create domain-specific debuggers with domain-specific views and operations. With the exception of the moldable debugger framework, all these solutions are dedicated tools and the developer is left alone to benefit from them in an integrated way. 


	\paragraph{Research questions.}
	Our research question is:
	\begin{itemize}
		\item What is a debugger API that is powerful and versatile enough to allow developers to perform many debugging tasks that would normally require specific debugging tools or tedious manual operations?
	\end{itemize}

In this paper we present \sindarin\footnote{
This work was supported by Ministry of Higher Education and Research, Nord-Pas de Calais Regional Council, CPER Nord-Pas de Calais/FEDER DATA Advanced data science and technologies 2015-2020.
The work is supported by I-Site ERC-Generator Multi project 2018-2022. We gratefully acknowledge the financial support of the M\'etropole Europ\'eenne de Lille.
}, a scriptable on-line debugger API for Pharo\footnote{Pharo  is a pure object-oriented dynamically typed programming language inspired by Smalltalk - \url{http://www.pharo.org}.}~\cite{Duca17a}, and we use it to reproduce several debugging scenarios and advanced breakpoints from the literature.
To do this, \sindarin exposes stepping and introspection operations not always appearing in mainstream debuggers such as VisualStudio or GDB.
In addition, it simplifies the creation of personalized debugging scripts by providing AST mappings, thus also proposing different stepping granularity over the debugging session.
Finally \sindarin facilitates object reachability to use object-centric debugging~\cite{Ress12a,Cost18a}.

	\paragraph{Contributions.}
	The contributions of this paper are:
	\begin{itemize}
		\item The identification of three main requirements that a debugger API must satisfy to allow developers to solve many debugging scenarios from the literature: step granularity at the expression level, full access to execution contexts~(reading and modifying) and a bytecode-to-AST mapping.
		\item The design of a debugger API through which developers can express several debugging scenarios and breakpoints from the literature, that would normally require specific tools.
	\end{itemize}

\paragraph{Outline.}
The paper is structured as follows: Section~\ref{overview} presents an overview of the \sindarin API and a usage example .
Section~\ref{scenarios} presents debugging scenarios from the literature and how they are solved with compact \sindarin scripts.
Section~\ref{sec:ExpressingAdvancedBreakpoints} shows how to use \sindarin to express advanced breakpoints from the literature.
Section~\ref{sec:ObjectCentricDebugging} shows how \sindarin facilitates object-centric debugging.
Section~\ref{sec:EvaluationAndRelatedWork} analyses the requirements of debugging scenarios and compares them to mainstream debugger APIs.
Section~\ref{sec:Implementation} gives a glance at the implementation of \sindarin in the Pharo debugger and its integration in the Integrated Development Environment. It also explains how \sindarin could be integrated in mainstream software development and discusses its limitations
Section~\ref{sec:RelatedWork} exposes the related work.
Finally, section~\ref{sec:discussion} touches upon other aspects around \sindarin, like self debugging or the generation of persistent artifacts from \sindarin scripts.


\vspace{-0.2cm}
\section{\sindarin's Overview}\label{overview}

\sindarin is a scripting API for the Pharo debugger, implemented as an internal DSL.
In this section we introduce \sindarin through an example making use of several of its features.
We then document an extract of \sindarin's API, as it is used in the rest of the paper to illustrate more advanced debugging scenarios.

\subsection{Sindarin by Example}

To illustrate \sindarin, let's consider a program that operates on several files inside a directory.
After some seconds running, the program tries to re-open an already open file and fails.
The developer would like the program to stop when a file is opened for the second time and know for \emph{this specific file} how and when it was opened the first time.
The \sindarin script shown in Listing~\ref{fig:sindarin-illustrated} solves this problem\footnote{For readers unfamiliar with the Pharo/Smalltalk syntax:
	\begin{itemize}
		\item The message-send notation uses spaces instead of dots, and there are no parenthesis to specify arguments: \ct{dbg currentNode} is equivalent to \ct{dbg.currentNode()}
		\item Arguments are specified by colons: \eg\\\ct{dbg messageReceiver isKindOf: File} \\is equivalent to\\ \ct{dbg.messageReceiver().\-isKindOf(File)}
		\item Square brackets \ct{[ ]} delimit lexical closures.
	\end{itemize}
}.
It~(a) first detects when a file is opened and records the stack when this happens, then~(b) it steps until an already open file is opened again.

\begin{lstlisting}[xleftmargin=5mm, label=fig:sindarin-illustrated, caption=\textbf{\Sindarin illustrated.} Step until a file is open twice and return the stack that was saved when it was first opened.]
result := nil
finished := false.
stackDictionary := Dictionary new.
[ finished ] whileFalse: [
(dbg currentNode isMessageNode
and: [(dbg messageReceiver isKindOf: File)
and: [ dbg messageSelector = #open ]])
ifTrue: [
result := stackDictionary
at: dbg messageReceiver
ifPresent: [ finished := true ]
ifAbsentPut: [ dbg stack copy ]].
dbg step ]
\end{lstlisting}
\vspace{-0.2cm}

This script continuously steps the execution~(lines 4 and 13).
After each step, it checks if the execution is about to send a message~(line 5) to an instance of the \ct{File} class~(line 6).
If the message is \ct{\#open}~(line 7), it checks in a dictionary whether the file has already been opened~(line 9-10).
If the file is being opened for the first time, an entry is added in the dictionary with a copy of the current execution stack as value~(line 12).
If there is already an entry~(line 11) the stack of interest is stored in \ct{result} and the script terminates.

This example illustrates several aspects of the API:
\begin{description}
	\item[Step Operations.] \sindarin commands the execution \\ through a \ct{step} method performing a step-into~(line~\(13\)).
	\item[Stack frame access.] The stack frame is queried multiple times during the script to get the receiver of a message before it is executed~(lines \(6,10\)) and to obtain the current stack trace~(line \(12\)).
	\item[AST Mapping.] \sindarin's \ct{currentNode} method maps the current program counter to the corresponding AST node, allowing high-level queries on the execution (lines~\(5,7\)).
	\item[Object-reachability.] All objects created and used during the execution are reachable by the \sindarin script, easing the construction of object-centric debugging scripts~(lines \(6,10\)).
\end{description}
\sindarin scripts are stateful: they can define and use variables.

\subsection{\sindarin's API}

\sindarin's internal DSL exposes among others:

	\begin{description}
		\item[Expression-level Step Operations.] Basic step operations like \ct{step into} and \ct{step over}. Unlike in most other debugging APIs, these operations work at the expression level rather than the line level.
		\item[Full Context Access.] Access to, and manipulation of the stack trace and its stack frames.
		\item[AST Mapping.] A bytecode-AST mapping, mapping the debugged program's program counter to the AST node currently executed.
		\item[Object-Centric Debugging Operations.]
		Scope break-\\ points to specific objects.
		\item[Breakpoint-related Operations.] To set, customize and remove breakpoints.
	\end{description}

\sindarin's API is further described in Table~\ref{tab:dbgAPI}.
The notations used in the table are as follows: \ct{dbg} stands for scriptable debugger, \ct{ctx} stands for a context~(a.k.a. stack frame), \ct{ast} stands for an AST node and \ct{bp} stands for breakpoint.

\begin{table*}
	\caption{The Sindarin debugging API}%
	\label{tab:dbgAPI}
	\begin{tabular}{l p{13cm}}
		\toprule
		\multicolumn{2}{ l }{\textbf{Stepping}} \\
		\midrule
		\ct{dbg step} & Executes the next instruction. If the instruction is a message-send, step inside it. \\
		\ct{dbg stepOver} & Executes over the next instruction. If the instruction is a message-send, step it until it returns. \\
		\ct{dbg stepUntil: aPredicate} & Steps the execution until the predicate is true. \\
		\ct{dbg skipWith: obj} & Skips the execution of the current instruction, and puts the object \ct{obj} on the execution stack. \\
		\ct{dbg skip} & Skips the execution of the current instruction, and puts \ct{nil} on the execution stack. \\
		\ct{dbg continue} & Steps the execution until a breakpoint is hit. Returns a reification of the breakpoint hit.\\
		
		\toprule
		\multicolumn{2}{ l }{\textbf{Stack Access}} \\
		\midrule
		\ct{dbg isExecutionFinished} & Returns whether the debugged execution is finished. \\
		\ct{dbg context} & Returns a reification of the current stack-frame. \\
		\ct{dbg stack} & Returns a list of context objects representing the current call stack. \\
		\ct{ctx pc} & Returns the current program counter of the given context. \\
		\ct{ctx sender} & Returns the sender context of the given context. \\
		\ct{ctx receiver} & Returns the receiver of the given context. \\
		\ct{ctx selector} & Returns the selector of the given context. \\
		\ct{ctx method} & Returns the method of the given context. \\
		\ct{ctx arguments} & Returns the arguments of the given context. \\
		\ct{ctx temporaries} & Returns the temporary variable of the given context. \\
		
		\toprule
		\multicolumn{2}{ l }{\textbf{Stack Modification}} \\
		\midrule
		\ct{ctx push: aValue} & Pushes aValue into the stack-frame's value stack.  \\
		\ct{ctx pop} & Pops a value from the stack-frame's value stack and returns it. \\
		
		\toprule
		\multicolumn{2}{ l }{\textbf{AST and AST Mapping}} \\
		\midrule
		\ct{dbg currentNode} & Returns the node that corresponds to the current program counter.\\
		\ct{ast accept: visitor} & Visits the current node using a visitor pattern. \\
		\ct{ast is*Node} & Returns \ct{true} if the receiver is a node of the specified kind~(for example: \ct{ast isMessageNode}). \\
		\toprule
		\multicolumn{2}{ l }{\textbf{Object-Centric Debugging}} \\
		\midrule
		\ct{dbg haltOnCall: obj} & Breaks next time the object \ct{obj} receives any message.\\
		\ct{dbg haltOnCall: obj for: m} & Breaks next time the object \ct{obj} receives the message \ct{m}.\\
		\ct{dbg haltOnWrite: obj } & Breaks next time any instance variable of the object \ct{obj} is written.\\
		\ct{dbg haltOnWrite: obj field: iv} & Breaks next time the instance variable \ct{iv} of the object \ct{obj} is written.\\
		\toprule
		\multicolumn{2}{ l }{\textbf{Breakpoints}} \\
		\midrule
		\ct{dbg setBreakpoint} & Sets a breakpoint on the current node, returns an object reifying the breakpoint.\\
		\ct{dbg setBreakpointOn: T} & Sets a breakpoint on \ct{T}~(a node or a compiled method), returns an object reifying the breakpoint.\\
		\ct{bp whenHit: aBlock} & Defines a sequence of debugging operations to perform when a breakpoint \ct{bp} is hit.\\
		\ct{bp remove} & Removes the breakpoint \ct{bp}.\\
		\ct{bp once} & Configures the Breakpoint \ct{bp} to remove itself the next time it is hit. Returns \ct{bp}.\\
		\toprule
		\multicolumn{2}{ l }{\textbf{Stack Access Helpers}} \\
		\midrule
		\ct{dbg receiver} & Returns the receiver of the current stack-frame. \\
		\ct{dbg selector} & Returns the selector of the current stack-frame. \\
		\ct{dbg method} & Returns the method of the current stack-frame. \\
		\ct{dbg arguments} & Returns the arguments of the current stack-frame. \\
		\ct{dbg temporaries} & Returns the temporary variables of the current stack-frame. \\
		\ct{dbg messageReceiver} & Returns the receiver of the message about to be sent, if the current node is a message node.\\
		\ct{dbg messageSelector } & Returns the selector of the message about to be sent, if the current node is a message node.\\
		\ct{dbg messageArguments} & Returns the arguments of the message about to be sent, if the current node is a message node.\\
		\ct{dbg assignmentValue } & Returns the value about to be assigned, if the current node is an assignment node.\\
		\ct{dbg assignmentVariableName} & Returns the variable name about to be assigned to, if the current node is an assignment node.\\
		\bottomrule
	\end{tabular}
\end{table*}

\section{Solving Debugging Scenarios with \sindarin}\label{scenarios}
In this section we present debugging scenarios frequently encountered by developers in the literature and we show how they can be solved with \sindarin.

\subsection{Monitoring Assignments to a Variable}
\label{scenario1}
In this scenario described originally in~\cite{Haih13a}, a developer spots that the \ct{foo} instance variable of the class \ct{Bar} has, at some point of the execution, the unexpected value of 42.
To find the cause of the bug, she wants to know which assignments store 42 into the variable, for what she needs to manually search all writes to that variable and set a conditional breakpoint on them.

\noindent
Fig.~\ref{lst:SindarinScriptScenario1} shows a \sindarin script solving this problem.
This script steps the execution until an assignment with value \ct{42}~(line 3) is about to be performed on the \ct{\#foo} variable~(line 4) on an instance of the \ct{Bar} class~(line 5).
The assignment is detected in line 2 using AST comparison.
This condition makes sure the script does not catch assignments to instance variables of the same name but from another class.
With this script, the developer can quickly find the unexpected assignment and proceed from there to find the cause of the bug.

\vspace{0.5cm}
\begin{lstlisting}[xleftmargin=5mm, label=lst:SindarinScriptScenario1, caption=Stopping on specific assignments with \sindarin.]
dbg stepUntil: [
dbg currentNode isAssignment
and: [ (dbg assignmentValue == 42)
and: [ (dbg assignmentVariableName = #foo)
and: [ dbg receiver isKindOf: Bar ]]]
]
\end{lstlisting}
\vspace{-0.3cm}

\subsection{Stopping Before an Exception}%
\label{scenario:exception}
When debugging, developers often want to stop the execution just before a certain situation occurs, for example a specific exception being raised.
To raise an exception in Pharo, the message \ct{signal} must be sent to a subinstance of the class \ct{Exception}, so this is what the script will look for.

Using \sindarin the developer can write the script shown in Listing~\ref{script:stepUntilAboutToSignalException}, that steps the execution until it is about to raise an exception.

\begin{lstlisting}[xleftmargin=5mm,label=script:stepUntilAboutToSignalException, caption=\Sindarin script to step until an exception is about to be signalled.]
dbg stepUntil: [
dbg currentNode isMessage
and: [ (dbg messageSelector = #signal)
and: [ dbg messageReceiver isKindOf: Exception ]]]
\end{lstlisting}

This scenario is very close to the one proposed by Haihan \etal where a developer observes a \ct{NullPointerException} in a chain of dereferences~\cite{Haih13a}, as illustrated by the following Java expression \ct{total.getObjects().addAll(current.getObjects())}.

In Java, when the exception is raised, the context where the problem happened and its intermediate expressions is lost: only the line number where the error was raised is available.
If the developer wants to find the problematic sub-expression, she needs to rewrite the expression with one statement per line, or use a complex chaining of step into operations.
However, even if the two problematics are, somehow, equivalent~(a specific exception is raised), this time the developer wants to stop before the call is performed, and not before the exception is sent.

A \sindarin script solving this problem is illustrated in Listing~\ref{lst:SindarinScriptScenario3}.
This script steps the execution until a message is about to be sent to a receiver with value \ct{nil}.
When the script finishes, the debugger is in the state desired by the developer: the execution is suspended in the problematic sub-expression.


\begin{lstlisting}[xleftmargin=5mm,label=lst:SindarinScriptScenario3, caption=Catching message-sends to \ct{nil}.]
dbg stepUntil: [
dbg currentNode isMessage
and: [ dbg messageReceiver = nil ]]
\end{lstlisting}

\subsection{Placing Breakpoints on a Family of Methods}
In this scenario, the execution opens the same file multiple times, through different file-opening methods whose names conform to a given regex pattern~(\eg \ct{'.*open.*File.*'}).
The developer wants to set breakpoints in all the methods that call one of these file-opening methods to open the given file~(\eg \ct{'myFile.txt'}).
With \sindarin, the developer writes the script shown in Listing~\ref{lst:scenario5}.
This way of expressing a location in the code using an expression is close to \textit{pointcuts} definitions in \textit{Aspect-Oriented Programming}~\cite{Kicz97a}.

\begin{lstlisting}[xleftmargin=5mm,label=lst:scenario5, caption=Setting breakpoints on a family of methods.]
[ dbg isExecutionFinished ] whileFalse: [
(( '.*open.*File.*' match: dbg selector ) and:
[ (dbg context arguments at: 1) = 'myFile.txt' ])
ifTrue: [ dbg setBreakpointOn: dbg context sender method ].
dbg step ]
\end{lstlisting}

This script steps through the whole execution~(line 1).
When the selector of the current method matches the pattern~(line 2) and the argument of the current method is the file the developer is interested in~(line 3), the script places a breakpoint in the method that called the current method~(line 4).
Once all the breakpoints are set, the developer can launch again his program and start the debugging session knowing she will stop each time the file \ct{myFile.txt} is opened.


\section{Expressing Advanced Breakpoints}\label{pitons}\label{sec:ExpressingAdvancedBreakpoints}
In this section, we show how to use \sindarin to express more advanced breakpoints.

\subsection{Control-Flow Breakpoints}
In~\cite{Cher07c}, Chern and De Volder present a breakpoint-definition language to define breakpoints based on control-flow aspects of the execution.
Their motivational example is that a method \ct{ProjectBrowser>>\#trySaveAs:} that they want to debug is called by multiple other methods during the execution.
They want the execution to stop in this method, but not if it is called by \ct{ActionSaveProjectAs>>\#actionPerformed:} because they identified that situation to be uninteresting.

To express this, a developer using \sindarin writes the script shown in Listing~\ref{lst:SindarinScriptScenario6}.
This script steps the execution until the current method is the desired one~(line 2) and the method of the sender context is not the undesired one~(lines 3-4).

\begin{lstlisting}[xleftmargin=5mm,label=lst:SindarinScriptScenario6, caption=\sindarin script for a control-flow breakpoint]
dbg stepUntil: [
(dbg method = ProjectBrowser>>#trySaveAs:) and:
[ (dbg context sender method =
ActionSaveProjectAs>>#actionPerformed:) not ]]
\end{lstlisting}

This scenario is slightly more complicated if the \ct{ActionSaveProjectAs>>\#actionPerformed:} method is not the direct sender but is lower in the stack.
The developer wants to stop the execution in the \ct{ProjectBrowser>>\#trySaveAs:} method, but not if the \ct{ActionSaveProjectAs>>\#actionPerformed:} method is present \emph{anywhere in the stack}.
To solve this extended scenario, the developer replaces line 3 of the original script to look up the entire stack, as shown in Listing~\ref{lst:SindarinScriptScenario6v2}.

\begin{lstlisting}[xleftmargin=5mm,label=lst:SindarinScriptScenario6v2, caption=\sindarin script for a control-flow breakpoint (extended scenario).]
dbg stepUntil: [
(dbg method = ProjectBrowser>>#trySaveAs:) and:
[ (dbg stack anySatisfy: [:ctx | ctx method = ActionSaveProjectAs>>#actionPerformed:) not ]]
\end{lstlisting}

\subsection{Chaining Pitons}
\label{Pitons}
Some bugs are only reproducible deep inside an execution.
Long loops and recursive structures are examples of common programming patterns that harm debugging.
Debugging in these scenarios is not only time consuming, but also requires patience and discipline from the developer.
On the one hand, manual stepping is error prone: an extra step can go over the cause of the bug, invalidating the debug session and forcing the developer to restart the debug session.
On the other hand, a breakpoint inside a complex computation could be triggered hundreds of times before the bug actually appears.

Pitons\footnote{Pitons in alpinism are sticks placed one by one to secure the path of an alpinist.
	An alpinist carefully moves its secure rope from one piton to the next one.}, also called stateful breakpoints~\cite{Bodd17a}, are a sequence of breakpoints that trigger only if they are activated in the right order.
That is, given a sequence of breakpoints $(b_1, b_2 ... b_n)$, $b_n$ will only trigger if $b_{n-1}$ already triggered, and $b_{n-1}$ will trigger only if $b_{n-2}$ already triggered, and so on.

Building such a breakpoint with \sindarin boils down to a sequence of instructions to step the execution until the pitons are reached.
Listing~\ref{lst:PitonsUsingStepUntil} shows a \sindarin script passing through pitons \ct{method1}, \ct{method2},... up to \ct{methodN}.

\begin{lstlisting}[xleftmargin=5mm,label=lst:PitonsUsingStepUntil, caption=\sindarin script stepping through pitons.]
dbg stepUntil: [ dbg method = ClassA>>#method1 ].
dbg stepUntil: [ dbg method = ClassB>>#method2 ].
...
dbg stepUntil: [ dbg method = ClassC>>#methodN ]
\end{lstlisting}

This script can be further enhanced by extracting common behaviour into a new stepping function that receives as argument a collection of piton methods~(Listing~\ref{lst:PitonsUsingStepUntil_refactored}).
The stepping function iterates over the collection of pitons and steps until it has found all of them during the execution.

\begin{lstlisting}[xleftmargin=5mm,label=lst:PitonsUsingStepUntil_refactored, caption=Refactored \sindarin script stepping through pitons.]
self stepThroughPitons: {ClassA>>#method1. ClassB>>#method2. ... ClassC>>#methodN}.

MyScript >> stepThroughPitons: anArray
anArray do: [:aMethod |
dbg stepUntil: [ dbg method = aMethod ]]
\end{lstlisting}

\subsection{Divergence Breakpoints}\label{sec:DebuggingStackTransformWithSindarin}
Introducing a minor change in the source code can cause a program to behave very differently.
When facing this situation, developers are left to manually step multiple executions with and without the changes, and compare them manually~\cite{Zell02a}.

\paragraph{The Pillar bug Scenario.} Pillar\footnote{https://github.com/pillar-markup/pillar} is an open-source document generator.
It takes a source document written in a markup syntax, and generates an output document~(\eg a pdf).
The Pillar test suite is made up of more than 3000 unit tests.
One of the tests failed after developers introduced an instance variable and its accessor in a Pillar class named \ct{Configuration}~\cite{Dupr17a, Cost18b}.
This bug was difficult to solve because the symptoms of the bug~(a failing test) had no relation with the modification~(introducing an accessor) and its supposed impact.

\paragraph{Finding Execution Divergence with \sindarin.}
To solve this scenario, we command two executions side-by-side and step them until they diverge \ie until both programs stack-traces differ.
First, we copied the \ct{Configuration} class and the failing test using this class.
One test~(\ct{TestOriginalClass}) executes the original behavior with the original class.
This test passes.
The second test~(\ct{TestCopyClass}) uses the copy class, in which we added an instance variable and its accessor method named \ct{\#disabledPhases}.
This test fails.
Listing~\ref{echo-script} shows  a script that commands the executions of the two tests until they step in different methods.
Lines 1-2, we create the two debug sessions for the two tests.
Line 3-4, we start a stepping loop, in which the two sessions are stepped side-by-side until the current method of both executions are different.

\begin{lstlisting}[xleftmargin=5mm,label=echo-script, caption=Script to debug the two executions side-by-side.]
dbg1 := ScriptableDebugger debug: [	TestOriginalClass run ].
dbg2 := ScriptableDebugger debug: [ TestCopyClass run ].
[ dbg1 method = dbg2 method ]
whileFalse:[	dbg1 step. dbg2 step ]
\end{lstlisting}

\paragraph{Analysis.}
At first, one execution uses an instance of the original \ct{Configuration} class, and the second execution uses an instance of its modified counterpart, with the initialized new instance variable and its accessor.
At some point of the execution, those instances are composed with new uninitialized instances of their own class, named \textit{sub-configuration}.
When a sub-configuration, instance of the modified \ct{Configuration} class, receives the \ct{\#disabledPhases} message, it answers an uninitialized value.
When a sub-configuration, instance of the original \ct{Configuration} class, receives the same message, it raises a \ct{\#doesNotUnderstand:} exception, which starts a hidden developer-defined lookup to retrieve the value from the original \ct{Configuration} instance.
We are immediately able to see where the two executions diverge, and notice that adding an accessor prevents the hidden lookup to start, thus producing our bug.


\subsection{Domain Specific Breakpoints}

Generic debugging operations such as \ct{step} or \ct{stepOver} are useful because they are applicable to executions of many different domains.
However, when using generic stepping operations, debugging high-level libraries and internal DSLs such as Roassal~\cite{Berg16c} or Glamour becomes a repetitive and error prone task. Indeed, the steps are too small and go through a lot of places that the developer knows to be uninteresting.
Let's take for example the internal iterators of the Pharo Collections library such as \ct{\#collect:}\footnote{In Pharo, the \ct{\#collect:} method implements the well-known \ct{map} function from functional programming.}, whose code is shown in Listing~\ref{collect}.
This iterator receives a block closure as argument, executes it over each element of the collection and returns a new collection with the collected elements.
This iterator is defined using the \ct{\#do:} iterator, which applies a block closure to every element of a collection. The \ct{\#do:} iterator is implemented differently for each collection.
When encountering a call to \ct{\#collect:}, the developer usually does not want to step inside the implementation of \ct{\#collect:}~(and even less inside the one of \ct{\#do:}).
Instead, she would like to jump through the Collection library's code to when her block closure is executed on the elements of the collection.

\begin{lstlisting}[xleftmargin=5mm,label=collect, caption=The \ct{\#collect:} internal iterator from Pharo collections., mathescape=true]
"Usage example"
self bigCollection collect: [ :each | each double ].

"Collect code"
Collection >> collect: aBlock
| newCollection |
newCollection := self species new.
self do: [:each | newCollection add: (aBlock value: each)].
$\hat{}$ newCollection
\end{lstlisting}

Listing~\ref{stepToNextIteration} illustrates how the developer writes a script that steps until the 173$^{th}$ iteration of a \ct{\#collect:} on a large collection.
Line 1 steps inside the \ct{\#collect:}.
Line 2 captures the first argument of the invoked method: the block closure.
We then call repeatedly a custom stepping function \ct{\#stepToNextIteration} that will step until the next invocation of the block.
This stepping detects the invocation of the block when the block is used as method.
Also, it recognizes a \emph{new invocation} as a context it has not seen before.

\begin{lstlisting}[xleftmargin=5mm,label=stepToNextIteration, caption=The \ct{\#collect:} internal iterator from Pharo collections.]
dbg stepUntil: [ dbg method = (Collection >> #collect:) ].
blockClosure := dbg arguments first.

"step to the third iteration"
173 timesRepeat: [ self stepToNextIteration ]

MyScript >> stepToNextIteration
dbg stepUntil: [ :ctx | lastCtx ~~ ctx and: [ ctx method = blockClosure ] ].
lastCtx := dbg currentContext
\end{lstlisting}

\section{Easing Object-Centric Debugging}
\label{sec:ObjectCentricDebugging}
Object-centric debugging focuses on debugging operations at the level of objects rather than static method representations and execution stacks~\cite{Ress12a}. For example, this includes stopping the execution when a \emph{specific} object receives a message, or when the state of this particular object is accessed or modified.
%

The first step to debug an object is to find that object but object-centric debuggers do not provide dedicated means to find objects. We first give an example of application for object-centric debugging, 
then, we show how scripting the debugger with \sindarin eases acquiring objects for object-centric debugging.

\subsection{Why Object-centric Debugging: an Illustration}

Imagine \textit{AtomViewer}, an incarnation of the legendary Self~\cite{Holz90a} graphical application, displaying lots of random small shapes called \textit{atoms}. Such atoms can be of different shapes~(squares, dots, circles, stars, etc.). Listing~\ref{atom-example} shows the method called by the display loop for each atom:

\begin{lstlisting}[xleftmargin=5mm,caption=How to debug atom drawers?, label=atom-example]
AtomViewer >> displayAtom: anAtom
anAtom renderWith: self randomAtomDrawer
CircleAtom >> renderWith: anAtomDrawer
anAtomDrawer renderCircle: self
TorusAtom >> renderWith: anAtomDrawer
anAtomDrawer renderTorus: self
SphereAtom >> renderWith: anAtomDrawer
anAtomDrawer renderSphere: self
\end{lstlisting}

Each atom is instance of the \ct{Atom} class. Every time the display is refreshed, each atom is redrawn through the \ct{displayAtom:} method. It takes the atom instance to display, and passes a \emph{random} drawer object to it~(line 2). This drawer knows how to display atoms, as atoms have different shapes. Each drawer uses specific display options to draw the atoms' shapes~(like motion blur, glow, blend...). To that end, drawers implement different display methods, and know which method to call depending on the display options.

Imagine a bug that sometimes occurs when circular atoms that have a radius higher than 50 pixels are rendered.
It is difficult to debug since we do not know which drawer is erratic, and drawers are randomly chosen at display time.
In addition, circular atoms~(spheres, torus, circles...) all call a different rendering behavior from drawers~(lines 4, 6, 8).
To avoid putting conditional breakpoints everywhere, we want to put an object-centric breakpoint on the drawer object.
We also want to set this breakpoint before the drawer is passed to an atom with a problematic radius, so that the drawer halts each time it receives a rendering message.
However, we cannot easily access the drawer object. It is returned by a call to the \ct{randomAtomDrawer} method, used immediately as a parameter to the \ct{acceptVisitor:} message, then discarded and garbage collected.


\subsection{Easing Object-centric Debugging with \sindarin}
\label{sec:capture}
%
%

%



With \sindarin, we control the execution when breakpoints are hit: we navigate the execution to reach the right context, from which we extract and capture objects of interest.
We apply object-centric debugging operations to those objects, then continue the execution.
It gives us a systematic and automatic means to capture objects for object-centric debugging.


%

In Listing~\ref{atom-example-2}, we capture the object returned by the \ct{self randomAtomDrawer} message in Listing~\ref{atom-example},
before it is used as parameter to the \ct{renderWith:} method.
To achieve this, we set a breakpoint on the \ct{displayAtom:} method~(line 1).
When the breakpoint is hit, we first do a \ct{stepOver}. It executes the \ct{self randomAtomDrawer} expression and puts the result on the value stack.
At this point, the debugger is about to send the \ct{renderWith:} message, with \ct{anAtom} as receiver, and the object that has been put on the stack as argument~(the atom drawer).
We use the execution introspection API to recover both objects~(lines 4-5) and check conditions~(line 6) before applying an object-centric breakpoint to the drawer~(line 8) and removing the original breakpoint~(line 9).

\begin{lstlisting}[xleftmargin=5mm,caption=Capture example: capturing an atom drawer., label=atom-example-2]
bpoint := dbg setBreakpointOn: AtomViewer>>#displayAtom:.
bpoint whenHit: [
dbg stepOver.
atom := dbg messageReceiver.
drawer :=  dbg messageArguments first.
(atom shape isCircular and: [atom shape radius > 50])
ifTrue: [
dbg haltOnCallTo: drawer.
bpoint remove].
dbg continue ]
\end{lstlisting}

%
%

\subsection{Replaying Objects}

\begin{figure}[htb]
	\includegraphics[width=\linewidth]{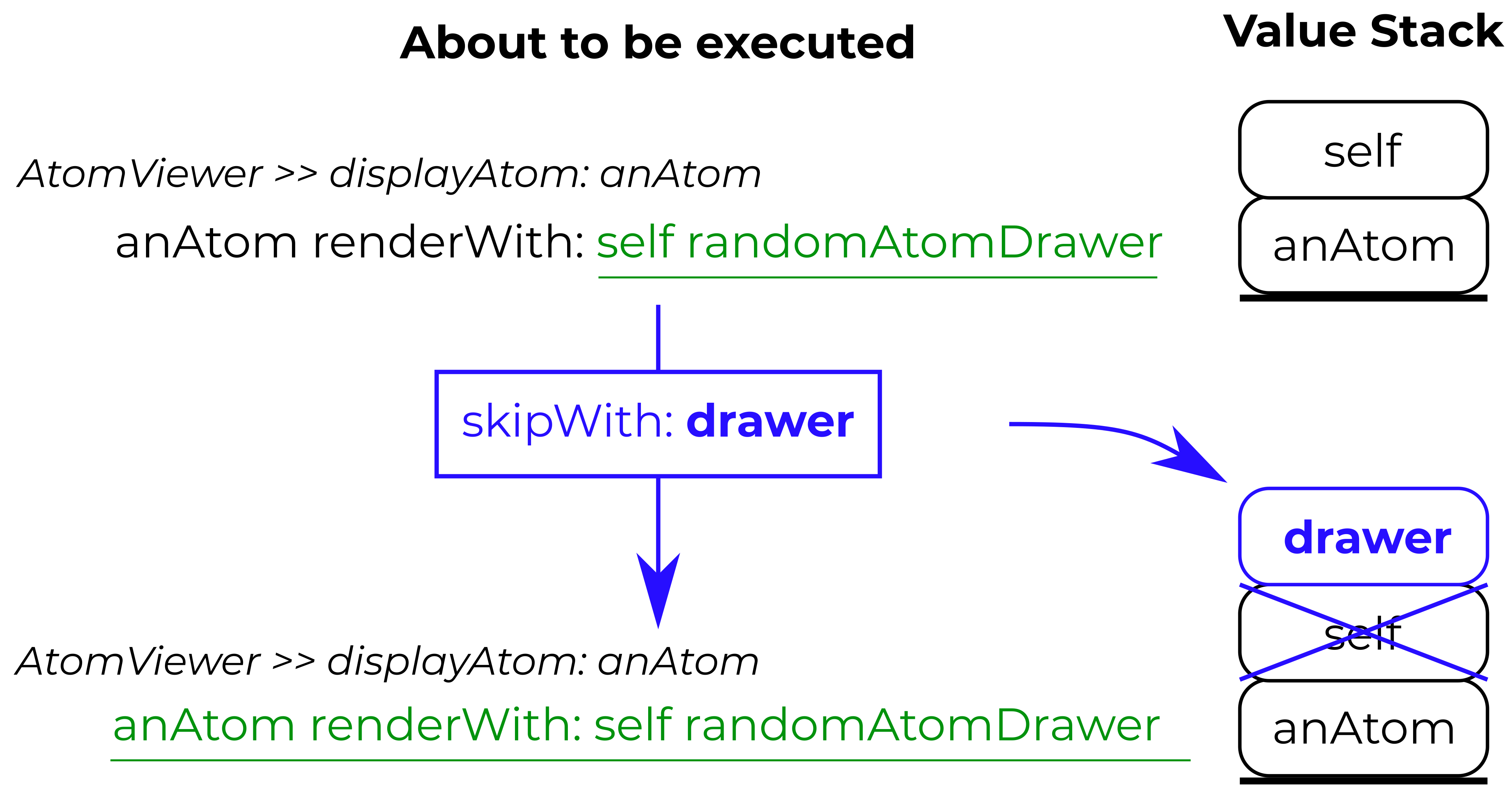}
	\caption{Removing non-determinism: \ct{skipWith: anObject} steps over a node about to be executed~(underlined) but without executing it, and pushes \ct{anObject} on the stack instead.}%
	\label{fig:replayDebugging}
\end{figure}

Bugs can be hard to reproduce when programs contain non-deterministic aspects.
For instance, in our example from Listing~\ref{atom-example}, the drawer object is chosen randomly.
To replay the bug, we re-inject the problematic drawer in the control flow using the \ct{skipWith:} operation~(illustrated in Figure~\ref{fig:replayDebugging}).

\vspace{-1mm}

We show in Listing~\ref{atom-replay} how to achieve this with \sindarin.
Each time we break, we capture the node whose execution creates a random drawer~(line 3), and the drawer it returns~(line 4-5).
If conditions for replay are satisfied~(line 6), we remove the original breakpoint and configure a replay breakpoint on the drawer node~(line 8).
When this replay breakpoint is hit~(line 9), we instruct the debugger to skip over the current node~(the drawer creation node) and use the captured drawer object instead~(line 10).

\begin{lstlisting}[xleftmargin=5mm,caption=Replaying an atom drawer after an exception, label=atom-replay]
bpoint := dbg setBreakpointOn: AtomViewer>>#displayAtom:.
bpoint whenHit: [
drawerNode := dbg currentNode.
dbg stepOver.
drawer :=  dbg messageArguments first.
"condition for replay" ifTrue: [
bpoint remove.
replayPoint := dbg setBreakpointOn: drawerNode.
replayPoint whenHit: [
dbg skipWith: drawer.
dbg continue]].
dbg continue]
\end{lstlisting}


\section{Evaluation}\label{sec:EvaluationAndRelatedWork}



We have already shown throughout the paper that \sindarin is a versatile scripting DSL that covers different debugging scenarios from the literature.
In this section we  \sindarin with respect to mainstream debugger APIs such as GDB or JDI.
We first identify the different features required by \sindarin to implement each debugging scenario described in this paper.
We then evaluate how those features are supported in several mainstream debuggers with debugger APIs.

\subsection{\Sindarin Debugger Requirements}

Each of the debugging scenarios used throughout this paper requires a different set of features from \sindarin.
We analyzed and categorized those requirements into three different axis:
\begin{description}
	\item[Step Granularity.] The granularity required for the step operation \ie the minimal amount of code that is executed between two step operations. In our debugging scenarios, we observed the need for two different granularities: a fine-grained \emph{expression} granularity and a coarse-grained \emph{method} granularity.
	\item[Context Access.] The access provided to the execution stack. We have categorized context access in two main operations. \emph{Basic access} refers to the access to the list of stack frames and their methods, receiver and argument objects, and their temporary variables. \emph{Full access} refers to the access to the intermediate values of expressions during the execution~(\emph{Stack access}), and the ability to modify the current stack to \eg pop or push a value into it~(\emph{Stack modification}).
	\item[AST Mappings.] Whether the scenario requires AST mappings or not to identify code of interest dynamically.
\end{description}

\begin{table*}[h]\caption{Feature Requirements of the Debugging Scenarios}%
	\label{tab:scenario_requirements}
	\begin{small}
		\begin{tabular}{ l l l c }
			\toprule
			\textbf{Scenario} & \textbf{Step Granularity} & \textbf{Context Access}  & \textbf{AST Mapping} \\
			\midrule
			Monitoring Assignments & Expression &  \multirow{2}{*}{Stack Access} & \multirow{2}{*}{\ding{51}} \\
			to a Variable & & & \\
			\midrule
			Stopping before & Expression & \multirow{2}{*}{Stack Access} & \multirow{2}{*}{\ding{51}} \\
			an Exception & & &\\
			\midrule
			Placing Breakpoints on& Method &  \multirow{2}{*}{Basic Access} & \multirow{2}{*}{\ding{55}} \\
			a Family of Methods& & & \\
			\midrule
			Chaining Pitons & Method &  Basic Access & \ding{55} \\
			\midrule
			Control-Flow Breakpoints & Method &  Basic Access & \ding{55}\\
			\midrule
			Divergence Breakpoints & Method &  Basic Access & \ding{55} \\
			\midrule
			DSL Stepping & Method &  Basic Access & \ding{55} \\
			\midrule
			Capturing Objects & Expression &  Stack Access & \ding{51} \\
			\midrule
			Replaying Objects & Expression &  Stack Modification & \ding{51} \\
			\bottomrule
	\end{tabular}\end{small}\label{table:sindarin-debug-API}
\end{table*}

Table \ref{tab:scenario_requirements} summarizes our analysis for each of our debugging scenarios.
We observed that several scenarios require expression step granularity to operate at the level of expressions instead of methods.
All scenarios requiring expression step granularity also require AST mappings to identify the interesting expressions, and stack access to access intermediate results of those expressions stored in the stack.
For instance, although it is possible, to some extent, to implement object-centric breakpoints with those debuggers, the object capture scenario is mandatory for their application. 
This scenario is one example requiring stack access to access intermediate results. 
Finally, replaying objects is the only scenario in our scenario collection to require stack modification. Replaying an object instead of an expression requires to skip the expression and push the object to replay in the stack.


\subsection{Comparison with Debugger APIs} \label{subsec:comparisondebuggerapi}

	Several debugger implementations provide an API or protocol to connect to it and drive the debugging of a program.
	Examples of these are GDB~\cite{gdb03} and Java's debugger interface through JDI and JVMTI~\cite{JDI,JVMTI}.
	In this section we take several mainstream debuggers and compare their APIs to \sindarin's, to analyze which debugging scenarios are applicable to them.
	
	The summary of our analysis is reflected in Table~\ref{table:ext-debug-API}.
	We identify five families of debugger APIs:~(1) Unmanaged-code debuggers~(UCD) covers debuggers for non-managed runtimes, for which we analyzed GDB, LLDB, WinDBG,~(2) Python's basic debugger functions~(BDB)~\cite{BDB,Coet15a},~(3) NodeJS debugging API,~(4) Java's debugger interface through JDI and JVMTI~\cite{JDI,JVMTI} and~(5) Microsoft's CLR EnvDTE debugger interface.

\begin{table*}[]
	\caption{Mainstream Debugging APIs.}
	\begin{tabular}{lcccccc}
		\textbf{Feature \textbackslash~API} & UCD$^{1}$ & BDB & NodeJS &  JDI & CLR$^{4}$   & \sindarin\\
		Step Granularity & line$^{2}$ & expression & line & line & line &expression\\
		Context Access & full$^{3}$  & basic & basic  & basic & basic &full\\
		AST Mapping & \ding{55} & \ding{55} & \ding{55} & \ding{55} & \ding{55} &\textbf{\ding{51}}\\
	\end{tabular}%
	\label{table:ext-debug-API}
	\begin{footnotesize}
		~\\
		~\\
		(1)  Unmanaged Code Debuggers: GDB, WINDBG, LLDB.\\
		(2) A line stepping operation at the machine instruction level is available.\\
		It matches the source code, but developers must know when to start and\\
		stop stepping to achieve an expression level granularity, as many assembly\\ instructions/bytecodes can be related to a single instruction in source code.\\
		(3)  Possible by reading and writing directly into the memory of the program.\\
		(4)  Common Language Runtime debuggers~( C\#, J\#, .NET).\\
	\end{footnotesize}
\end{table*}

	From our analysis we observe that most debuggers provide a rather coarse grained granularity for stepping of an entire line of code, which prevents a straight forward implementation of the scenarios discussed in this paper.
	Indeed, this granularity does not allow developers to differentiate between different expressions~(and sub-expressions) within a single line.
	Unmanaged code debuggers, on the other hand, provide by default a \ct{step} operation that operates at the line level while a second stepping operation does it at the machine instruction level.

	Another point of divergence is the fact that not all debugger APIs allow full stack manipulation, preventing advanced scenarios such as object-replay. We consider that unmanaged code debuggers support full stack manipulation since they can modify the execution stack and the objects pointed by it as any other memory region.

	Finally, no mainstream debugger provides high level mappings between instructions and AST such as \sindarin's AST mappings.

\vspace{-0.2cm}
\section{Implementation}\label{sec:Implementation}
We implemented \sindarin for the Pharo programming language.
In this section we briefly describe the infrastructure of our implementation, how it integrates within the Integrated Development Environment, how it could be integrated into mainstream software development, as well as its limitations.

\begin{figure}[h]
	\includegraphics[width=\linewidth]{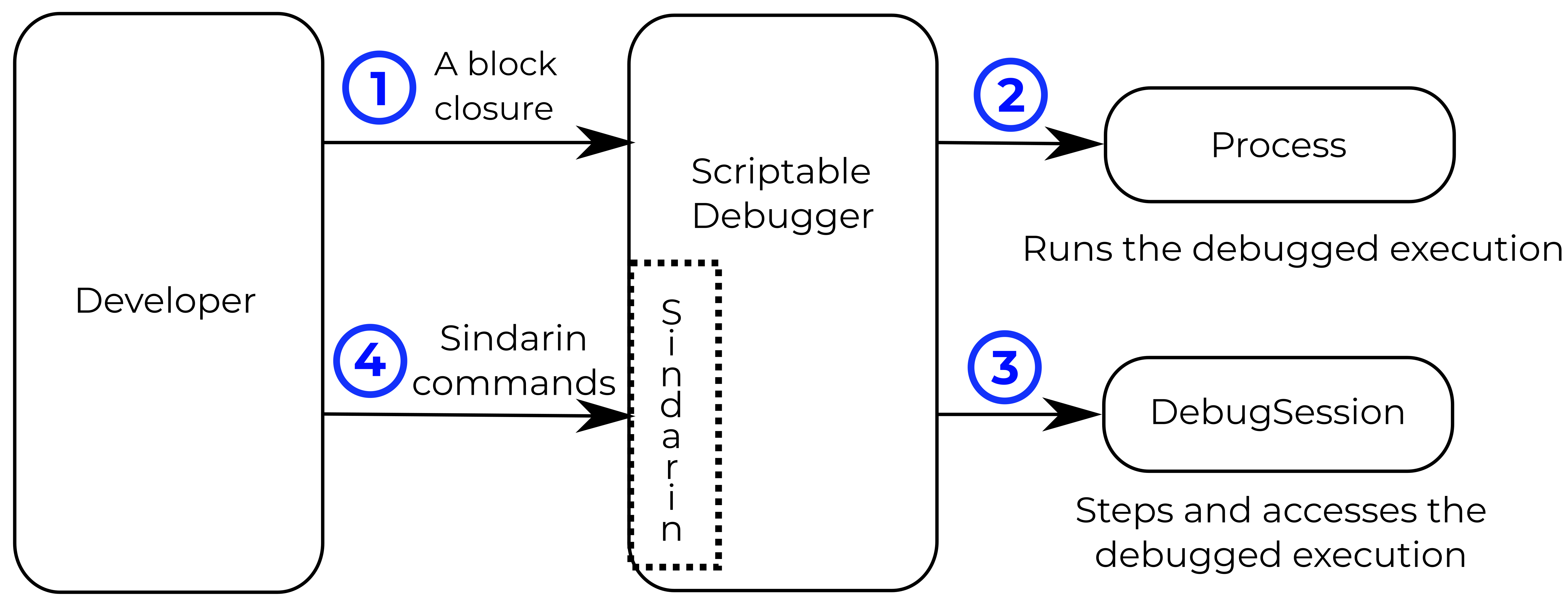}
	\caption{Creation process of a scriptable debugger instance}
	\label{fig:SindarinInfrastructure}
\end{figure}
\vspace{-1mm}

\subsection{Infrastructure}

\paragraph{Starting a \sindarin session.}

\sindarin drives a scriptable debugger instance and configures it for a dedicated piece of code that will be executed.
Figure~\ref{fig:SindarinInfrastructure} shows the creation process of a scriptable debugger instance.
The developer first provides a block closure of the code she wants to debug~(1).
The scriptable debugger then creates a process to run the closure~(2), and a \ct{DebugSession} to step and access the execution being debugged~(3).
Finally, the developer sends commands written using the \sindarin API to the scriptable debugger to control and/or inspect the debugged execution~(4).

	\paragraph{Contexts.}
	The contexts exposed by our \sindarin implementation are of the same class as the ones returned by the Smalltalk \ct{thisContext} pseudo-variable. 

	\paragraph{Step Granularity.}
	Our implementation of \sindarin runs in Pharo. Pharo code is compiled into bytecode, that is interpreted by a virtual machine. Through reflection, Pharo programs can access the bytecode interpreter of the virtual machine and step it to progress a frozen execution. The \ct{dbg step} operation of \sindarin is implemented using this feature to provide a stepping granularity at the expression level.

	\paragraph{Context Access.}
	\sindarin provides full access to the contexts of the debugged execution through instances of the \ct{Context
	} class. Context objects are automatically reified by the Pharo virtual machine when needed for reflection purposes.

\paragraph{Mapping AST to bytecode.}
The Pharo compiler keeps a mapping between the AST and the bytecode, as well as a mapping between the AST and the source code.
By default, the granularity of each step is at the level of expressions. 
\sindarin makes use of this infrastructure to provide fine grained control over stepping, without losing the bytecode abstraction.
This granularity eases the ability to set breakpoints inside statements.
The direct access and handling of the AST also enables to express complex conditions for breakpoints.

	\paragraph{Breakpoints.}
	The imperative design of \sindarin means that users will often set breakpoints on the AST node the debugged execution is currently at. Typical implementation of breakpoints modify the bytecode of the method to insert the breakpoint behaviour. This approach does not fit the use-case of \sindarin, because modifying methods that are already on the call stack is hard~\cite{Teso17b}. To go around this limitation, the breakpoints set by \sindarin are "virtual" in the sense that no code modification is performed. Instead, \sindarin remembers the AST nodes on which breakpoints were set, and stops debugged execution that would step through these nodes.

\subsection{IDE Integration}
\begin{figure}[h]
	\includegraphics[width=\linewidth]{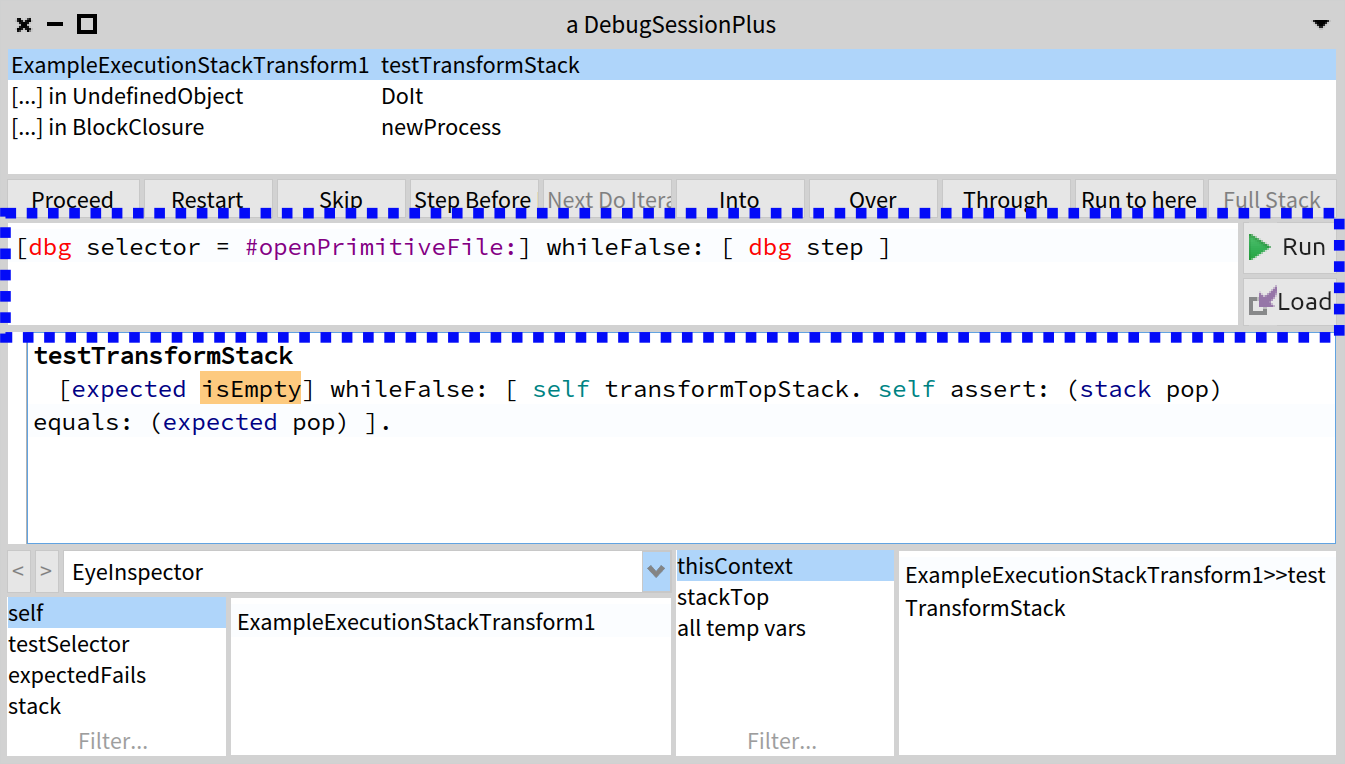}
	\caption{Command line implementing \sindarin integrated into the Pharo debugger}
	\label{fig:DebuggerCommandLine}
\end{figure}

We modified the Pharo debugger to make it scriptable through the \sindarin API. 
The developer writes \sindarin scripts in an editor or directly in the debugger~(Figure~\ref{fig:DebuggerCommandLine}). 
When scripts are run in the debugger, the \ct{dbg} variable is automatically bound to the current debugger. 

\vspace{-0.2cm}
\subsection{Integration into Mainstream Software Development}


	We highlighted a few existing debugger APIs from mainstream languages in section~\ref{subsec:comparisondebuggerapi}. These could be used as a basis to implement the \sindarin API in these languages. The three main challenges of such implementations are those outlined in section~\ref{subsec:comparisondebuggerapi}: \ct{step granularity}, \ct{context access} and \ct{AST mapping}.
	
	Implementing expression-level step granularity in the languages that do not natively support it can be achieved with automatic program rewriting~\cite{Kume19a}, to rewrite the expressions over multiple lines, or AOP weaving~\cite{Haih13a}.
	
	From Table~\ref{table:ext-debug-API}, UCD already provides full context access through direct memory access. For Python, Python code can interact in an indirect way with the Python virtual machine to access/modify the context stack of the execution. An alternative method is probably achievable for Java. Of note is that these solutions require significant engineering effort.
	
	AST mapping is not present in any of the debugging APIs we considered. It could maybe be achieved by modifying the compiler of the host language to keep the AST that is generated during the compilation, and build a binding between the bytecode generated by the compiler and this AST.


\vspace{-0.2cm}
\subsection{Limitations}\label{subsec:Limitations}
	A limitation of \sindarin is its low-level aspect. It offers a lot of possibilities, but lacks a higher-level layer that would offer a quick and expressive way for developers to express higher-level debugging operations. This limitation is mitigated by the ability to write and re-use debugging scripts.	Another limitation is that effectively using \sindarin to debug requires understanding and manipulating the concepts of contexts, AST nodes and stack.

\vspace{-0.2cm}
\section{Related work}\label{sec:RelatedWork}
We present in the following related work on debugger customization and means to control debugging sessions.
\vspace{-0.2cm}
\subsection{Domain Specific Debugging Operations}\label{subsec:custom-bps}
Kompos~\cite{Marr17a} is a concurrency-agnostic debugger protocol, which decouples the debugger from the concurrency models employed by the target application.
As a result, the underlying language run-time can define custom breakpoints, stepping operations, and execution events for each concurrency model it supports,
and a debugger can expose them without having to be specifically adapted.
With \sindarin, it is possible to place standard breakpoints and to step in concurrent code.
We plan to introduce additional support for concurrent debugging, including specific breakpoints like Kompos'.

The Moldable Debugger framework~\cite{Chis15c} allows developers to create domain-specific debuggers, by defining and combining domain-specific debugging operations and views.
The moldable debugger adapts itself to a domain by selecting at run-time the appropriate appropriate debugging operations and views.
Debugging operations executes the program until a debugging predicate is matched, or performs an action every time a debugging predicate is matched.
Debugging predicates are either primitive predicates like attribute reads or method calls, or combinations of these.
A key difference with \sindarin is the ``immediacy''.
With \sindarin, developers write a script on the fly to help with the particular bug under investigation.
By contrast, the moldable debugger framework is geared towards investing time into creating a debugger able to assist the debugging of a specific domain, rather than a specific bug.
Both approaches are not incompatible, we believe that \sindarin can be used to build moldable debuggers and to script their specialized parts.

Expositor~\cite{Phan13a}, inspired by the work on MzTake~\cite{Marc04b}, combines scripting and time-travel debugging. The fundamental abstraction provided by Expositor is the execution trace, which is a time-indexed sequence of program state snapshots. Developers can manipulate traces as if they were simple lists with operations such as map and filter. From that perspective Expositor has a conceptual interface close to the Pharo \ct{thisContext} pseudo-variable that reifies stack on demand.

Barr \etal describe a time-traveling debugger for JavaScript/Node.js~\cite{Barr16a}. In particular, they describe three time-travel operations: \emph{reverse-step}~(rs) to step to the previously executed line in the \emph{current} stack frame, \emph{reverse-step dynamic}~(rsd) to step back in time to the previously executed statement in the \emph{any} frame, and \emph{reverse to callback origin}~(rcbo): a more specific operation to step back from the currently executing callback to the point in time when the callback was registered. These three operations are good candidates to extend the \sindarin API with time-traveling capabilities.

CBD is a Control-flow Breakpoint Debugger~\cite{Cher07c}.
CBD uses a dynamic pointcut language to characterise control-flow breakpoints.
These breakpoints are conditions on the control-flow, through which they were reached.
As such, CBD leverages \emph{aspect-oriented programming} to assist debugging operations.
The difference between CBD and \sindarin is the versatility.
CBD is specialised in expressing control-flow breakpoints, and as a result expresses them in a concise fashion.
\sindarin allows developers to express the same breakpoints as CBD, and to combine these breakpoints with other features such as object-centric debugging.

Bugloo~\cite{Ciab03a} is a source level debugger for Scheme programs, that are compiled into JVM bytecode by the Bigloo compiler.
Bugloo traces methods that are entered, and the source code location where a given variable is read/written.
The debugging commands are recorded, which allows developers to replay their debugging sessions.
A command line is available for interactive debugger control.

Other solutions focus on exposing bugs through execution traces.
PTQL~\cite{Gold05a} is a query language, through which developers express what part of their program they want to trace.
Queries are automatically translated to instrumentation.
ange \etal~\cite{Lang95a} combines Prolog-like queries on program traces and visualization to understand program execution.
Working on traces implies a post-mortem approach, such works do not support controlling step by step execution of the debugged program, the precise queries of an execution moment nor object-centric facilities.

\vspace{-0.4cm}
\subsection{Querying Objects}
Finding objects of interest for debugging  is difficult.
Using traditional debugging tools, developers have to put breakpoints to halt the execution and manually select objects~\cite{Ress12a,Corr01a}.
This approach requires much interaction from developers and forbids any automation, \textit{e.g.} breakpoints in loops break the execution at each loop iteration.

Developers need systematic ways to obtain objects, for object-centric debugging or to provide more details about the debugged execution.
Fox~\cite{Pota04a} is a language to perform queries over the object graph of an execution snapshot.
Bugloo~\cite{Ciab03a} provides a heap inspector allows some access to the object graph of the execution.
\textit{Query-based debugging}~\cite{Lenc97a, Lenc99a} refers to debuggers which build collections of objects from user-defined queries, written in a dedicated query language.
\textit{Reactive object queries}~\cite{Lehm16a} are user-defined requests on a program, which results create and maintain dynamic collections of objects. Both solutions automatically search the object space to update their collections of objects.
However they cannot express conditions regarding the context from which an object is obtained~\cite{Cost18a}.
These solutions are good candidates for extending \sindarin.
\sindarin allows the developer to bring executions to points of interest, from where developers could perform queries over the object space.

\vspace{-0.2cm}
\subsection{Aspect-Oriented Programming}
	Aspect-oriented programming (AOP)~\cite{Kicz97a, Coly05a, Fabr12b} is a programming paradigm where the developer can add cross-cutting behavior (an "advice") to existing code without modifying the latter, by specifying which code is modified via a "pointcut" specification. AOP can be used for debugging purposes~\cite{Haih13a} and is able to solve some of the scenarios we described in this paper. AOP cannot directly tackle the Divergence Breakpoints scenario, as it requires two executions to be ran side-by-side. The Capturing Objects and Replaying Objects scenarios could be 
	implemented with AOP. This would however require additional work to define a model for storing and re-using objects across executions. AOP differs from \sindarin in the way debugging operations are expressed. \sindarin is imperative while AOP is declarative.
	The imperative paradigm is closer to the interactions developers have with mainstream debuggers: for example repeatedly clicking on the "step" button until the desired state is reached.
	We could use AOP to implement the \sindarin API in other languages.
	We leave as future work a comparison between the imperative (Sindarin) and declarative (AOP) styles in terms of learning curve and convenience for developers.





\vspace{-0.2cm}
\section{Discussion}\label{sec:discussion}
	In this section, we discuss a few points of note about \sindarin, namely the advantages of implementing it as an internal DSL, the ability and limits of debugging a \sindarin implementation with itself, and ideas about how \sindarin scripts written for debugging purposes could be leveraged as persistent artifacts.

\vspace{-0.2cm}
\subsection{Internal DSL}
	Our implementation of \sindarin is an internal DSL of Pharo. This brings advantages in terms of usability, as a developer debugging Pharo code will already be familiar with the general syntax of \sindarin scripts~(e.g. how to define a variable, block closures...) and have access to the standard libraries. In addition, this makes \sindarin directly usable within the Pharo IDE, with no additional integration work.
	
\vspace{-0.2cm}
\subsection{Self-debugging}
	
	Since \sindarin is an internal DSL, a debugging session using \sindarin is a program execution in the same host language. As such, \sindarin can be used to debug itself. This is in contrast to other debugging tools that cannot debug themselves. For example, Aspect-oriented programming is unable to debug itself in this fashion, as aspects cannot be added to other aspects~\cite{Haih13a}.
	A limit to the self-debugging ability of \sindarin is when a core feature is impacted, for example the \ct{step} operation itself. Indeed, if \ct{step} itself does not work, the buggy execution of \ct{step} cannot be stepped to be debugged. This limitations echoes that of Kansas~\cite{Smit97b}: a reflective system where developers interact with objects in a world. When a Kansas world is broken, another one is created from which the first world can be repaired. However, when a core feature like the ability to create new worlds is broken, Kansas cannot be debugged from within itself.
	
\vspace{-0.3cm}
\subsection{Artifact Generation}
	
	By debugging with \sindarin, a developer generates debugging scripts. These scripts constitute valuable artifacts that can be harvested and used for other purposes. For example, a script written to reach a certain point in the execution and check a property on a variable can be turned into a test. A script defining a particular domain-specific stepping operation~(like in our DSL Stepping scenario) can be kept to assist future developers in debugging the same section of the code. A script written to reach a point in the execution that is relevant for a given bug can be shared on the issue tracker. This allows other developers to immediately reach the relevant point without requiring a tedious textual description of which buttons to click in the debugger. These alternative, long-term uses for \sindarin scripts give additional value to the time spent by developers to write them during debugging sessions.

	Additionally, debugging scripts could be automatically generated from a developer's interaction with a mainstream debugger. For example the sequence of buttons she clicked.
\vspace{-0.2cm}
\section{Conclusion}
In this paper we presented \sindarin, a versatile API for scripting an online debugger.
\sindarin supports the definition of advanced stepping operations: It offers different interfaces such as AST or runtime stack introspection letting the developer use the abstractions she needs. \sindarin supports object-reachability in the sense that all the objects accessed during execution can be manipulated. This eases the construction of object-centric debugging scripts.
\sindarin will be shipped with Pharo 80 as a central part of the new debugging solution.

As a future work we will add support for concurrent programming debugging, back-in-time debugging, and provide ways to support the debugging based on multiple execution stacks. 
	In addition, we would like to perform a user-study with experimented developers to assess multiple characteristics of \sindarin like its learning-curve and how much it helps with debugging scenarios that are impractical to tackle without the support of specialized debugging tools. This study could also compare \sindarin to aspect-oriented-programming techniques on these two points to evaluate the difference between the imperative and declarative styles of debugging.

\bibliographystyle{abbrv}
\bibliography{otherspatched,rmodpatched}

\begin{thebibliography}{10}

\bibitem{Barr16a}
E.~Barr, M.~Marron, E.~Maurer, D.~Moseley, and G.~Seth.
\newblock Time-travel debugging for javascript/node.js.
\newblock In {\em Proceedings of the 2016 24th ACM SIGSOFT International
  Symposium}, pages 1003--1007, nov 2016.

\bibitem{Berg16c}
A.~Bergel.
\newblock {\em Agile Visualization}.
\newblock LULU Press, 2016.

\bibitem{Berg11h}
A.~Bergel, F.~Banados, R.~Robbes, and D.~R\"othlisberger.
\newblock Spy: A flexible code profiling framework.
\newblock {\em Journal of Computer Languages, Systems and Structures}, 38(1),
  Dec. 2011.

\bibitem{Bodd17a}
E.~Bodden.
\newblock Stateful breakpoints: A practical approach to defining parameterized
  runtime monitors.
\newblock In {\em ESEC/FSE'11}, 2011.

\bibitem{Cher07c}
R.~Chern and K.~De~Volder.
\newblock Debugging with control-flow breakpoints.
\newblock In {\em Proceedings of the 6th International Conference on
  Aspect-oriented Software Development}, AOSD '07. ACM, 2007.

\bibitem{Chis15c}
A.~Chis, M.~Denker, T.~Girba, and O.~Nierstrasz.
\newblock Practical domain-specific debuggers using the moldable debugger
  framework.
\newblock {\em Journal of Computer Languages, Systems and Structures},
  44:89--113, 2015.

\bibitem{Ciab03a}
D.~Ciabrini and M.~Serrano.
\newblock Bugloo: A source level debugger for scheme programs compiled into jvm
  bytecode.
\newblock In {\em Proceedings of the International Lisp Conference 2003}, oct
  2003.

\bibitem{Coet15a}
A.~L. Coetzee.
\newblock {\em Combining reverse debugging and live programming towards visual
  thinking in computer programming}.
\newblock PhD thesis, Stellenbosch University, 2015.

\bibitem{Coly05a}
A.~Colyer and C.~A.
\newblock Aspect-oriented programming with aspectj.
\newblock {\em IBM Systems Journal}, 44(2):301--308, 2005.

\bibitem{Corr01a}
C.~Corrodi.
\newblock Towards efficient object-centric debugging with declarative
  breakpoints.
\newblock In {\em SATToSE 2016}, 2016.

\bibitem{Cost18b}
S.~Costiou.
\newblock {\em {Unanticipated behavior adaptation : application to the
  debugging of running programs}}.
\newblock Theses, {Universit{\'e} de Bretagne occidentale - Brest}, Nov. 2018.

\bibitem{Cost18a}
S.~Costiou, M.~Kerboeuf, A.~Plantec, and M.~Denker.
\newblock Collectors.
\newblock In {\em {PX'18 - Programming Experience 2018}}, Companion of the 2nd
  International Conference on Art, Science, and Engineering of Programming,
  page~9, Nice, France, Apr. 2018. {ACM Press}.

\bibitem{Duca17a}
S.~Ducasse, D.~Zagidulin, N.~Hess, D.~C.~O. written~by A.~Black, S.~Ducasse,
  O.~Nierstrasz, D.~P. with D.~Cassou, and M.~Denker.
\newblock {\em Pharo by Example 5}.
\newblock Square Bracket Associates, 2017.

\bibitem{Dupr17a}
T.~Dupriez, G.~Polito, and S.~Ducasse.
\newblock Analysis and exploration for new generation debuggers.
\newblock In {\em Proceedings of the 12th Edition of the International Workshop
  on Smalltalk Technologies}, IWST '17, pages 5:1--5:6, New York, NY, USA,
  2017. ACM.

\bibitem{Fabr12b}
J.~Fabry and D.~Galdames.
\newblock Phantom: a modern aspect language for pharo smalltalk.
\newblock {\em Software: Practice and Experience}, 2012.

\bibitem{BDB}
P.~S. Foundation.
\newblock bdb - debugger framework.
\newblock \url{https://docs.python.org/3/library/bdb.html }, 2019.

\bibitem{Gold05a}
S.~Goldsmith, R.~O'Callahan, and A.~Aiken.
\newblock Relational queries over program traces.
\newblock In {\em Proceedings of Object-Oriented Programming, Systems,
  Languages, and Applications (OOPSLA'05)}, pages 385--402, New York, NY, USA,
  2005. ACM Press.

\bibitem{Holz90a}
U.~H{\"o}lzle, B.-W. Chang, C.~Chambers, and D.~Ungar.
\newblock {\em The {SELF} Manual}.
\newblock Computer Systems Laboratory of Stanford University, 1991.

\bibitem{JVMTI}
Sun microsystems, inc. {JVM} tool interface ({JVMTI}).
\newblock http://java.sun.com/j2se/1.5.0/docs/guide/jvmti/.

\bibitem{Kicz97a}
G.~Kiczales, J.~Lamping, A.~Mendhekar, C.~Maeda, C.~Lopes, J.-M. Loingtier, and
  J.~Irwin.
\newblock {Aspect-Oriented Programming}.
\newblock In M.~Aksit and S.~Matsuoka, editors, {\em Proceedings ECOOP '97},
  volume 1241 of {\em LNCS}, pages 220--242, Jyvaskyla, Finland, June 1997.
  Springer-Verlag.

\bibitem{Ko08a}
A.~J. Ko and B.~A. Myers.
\newblock Debugging reinvented: Asking and answering why and why not questions
  about program behavior.
\newblock In {\em Proceedings of the International Conference on Software
  Engineering, ICSE 08}, 2008.

\bibitem{Kume19a}
I.~Kume, E.~Shibayama, M.~Nakamura, and N.~Nitta.
\newblock Cutting java expressions into lines for detecting their evaluation at
  runtime.
\newblock In {\em Proceedings of the International Conference on Geoinformatics
  and Data Analysis}, ICGDA 2019, pages 37--46, New York, NY, USA, 2019. ACM.

\bibitem{Lang95a}
D.~Lange and Y.~Nakamura.
\newblock Interactive visualization of design patterns can help in framework
  understanding.
\newblock In {\em Proceedings ACM International Conference on Object-Oriented
  Programming Systems, Languages and Applications (OOPSLA'95)}, pages 342--357,
  New York NY, 1995. ACM Press.

\bibitem{Lehm16a}
S.~Lehmann, T.~Felgentreff, J.~Lincke, P.~Rein, and R.~Hirschfeld.
\newblock Reactive object queries.
\newblock In {\em Constrained and Reactive Objects Workshop (CROW)}, 2016.

\bibitem{Lenc97a}
R.~Lencevicius, U.~H{\"o}lzle, and A.~K. Singh.
\newblock Query-based debugging of object-oriented programs.
\newblock In {\em OOPSLA'97}, pages 304--317, 1997.

\bibitem{Lenc99a}
R.~Lencevicius, U.~H{\"o}lzle, and A.~K. Singh.
\newblock Dynamic query-based debugging.
\newblock In R.~Guerraoui, editor, {\em Proceedings of European Conference on
  Object-Oriented Programming (ECOOP'99)}, volume 1628 of {\em LNCS}, pages
  135--160, Lisbon, Portugal, June 1999. Springer-Verlag.

\bibitem{Lieb97a}
H.~Lieberman.
\newblock Introduction.
\newblock {\em Commun. ACM}, 40(4):26--29, Apr. 1997.

\bibitem{Marc04b}
G.~Marceau, G.~H. Cooper, S.~Krishnamurthi, and S.~P. Reiss.
\newblock A dataflow language for scriptable debugging.
\newblock In {\em IEEE International Conference on Automated Software
  Engineering (ASE'04)}, 2004.

\bibitem{Marr17a}
S.~Marr, C.~Lopez, D.~Aumayr, E.~Gonzalez~Boix, and H.~Mossenbock.
\newblock Kompos: A platform for debugging complex concurrent applications.
\newblock In {\em Programming'17}, pages 1--2, apr 2017.

\bibitem{JDI}
Oracle.
\newblock Java debug interface (jdi).
\newblock
  \url{http://docs.oracle.com/javase/7/docs/jdk/api/jpda/jdi/index.html}, 2013.

\bibitem{Pers17a}
M.~Perscheid, B.~Siegmund, M.~Taeumel, and R.~Hirschfeld.
\newblock Studying the advancement in debugging practice of professional
  software developers.
\newblock {\em Software Quality Journal}, 25(1):83--110, 2017.

\bibitem{Phan13a}
K.~Y. Phang, J.~S. Foster, and M.~Hicks.
\newblock Expositor: Scriptable time-travel debugging with first-class traces.
\newblock In {\em International Conference on Software Engineering (ICSE)},
  pages 352--361, may 2013.

\bibitem{Pota04a}
A.~Potanin, J.~Noble, and R.~Biddle.
\newblock Snapshot query-based debugging.
\newblock In {\em Proceedings of the Australian Software Engineering Conference
  (ASWEC'04)}, page 251. IEEE Computer Society, 2004.

\bibitem{Ress12a}
J.~Ressia, A.~Bergel, and O.~Nierstrasz.
\newblock Object-centric debugging.
\newblock In {\em Proceeding of the International Conference on Software
  Engineering}, 2012.

\bibitem{gdb03}
S.~S. Richard~Stallman, Roland~Pesch.
\newblock {\em Debugging with GDB}.
\newblock Gnu Press, 2003.

\bibitem{Smit97b}
R.~B. Smith, M.~Wolczko, and D.~Ungar.
\newblock From kansas to oz: collaborative debugging when a shared world
  breaks.
\newblock {\em Commun. ACM}, 40(4):72--78, Apr. 1997.

\bibitem{Spin18a}
D.~Spinellis.
\newblock Modern debugging: The art of finding a needle in a haystack.
\newblock {\em Commun. ACM}, 61(11):124--134, Oct. 2018.

\bibitem{Tass02a}
G.~Tassey.
\newblock The economic impacts of inadequate infrastructure for software
  testing.
\newblock {\em National Institute of Standards and Technology}, 2002.

\bibitem{Tell01a}
M.~Telles and Y.~Hsieh.
\newblock {\em The science of debugging}.
\newblock Coriolis Group Books, 2001.

\bibitem{Teso17b}
P.~Tesone, G.~Polito, L.~Fabresse, N.~Bouraqadi, and S.~Ducasse.
\newblock Dynamic software update from development to production.
\newblock {\em Journal of Object Technology}, 2018.

\bibitem{Vess86a}
I.~Vessey.
\newblock Expertise in debugging computer programs: An analysis of the content
  of verbal protocols.
\newblock {\em {IEEE} {Transactions} on {Systems}, {Man}, and {Cybernetics}},
  16, 1986.

\bibitem{Haih13a}
H.~Yin.
\newblock {\em Defusing the Debugging Scandal - Dedicated Debugging
  Technologies for Advanced Dispatching Languages}.
\newblock PhD thesis, University of Twente, Dec. 2013.

\bibitem{Zell09a}
A.~Zeller.
\newblock {\em Why programs fail: a guide to systematic debugging}.
\newblock Elsevier, 2009.

\bibitem{Zell02a}
A.~Zeller and R.~Hildebrandt.
\newblock Simplifying and isolating failure-inducing input.
\newblock {\em IEEE Transactions on Software Engineering}, SE-28(2):183--200,
  Feb. 2002.

\bibitem{Zhan10a}
C.~Zhang, D.~Yan, J.~Zhao, C.~Yuting, and S.~Yang.
\newblock Bpgen: An automated breakpoint generator for debugging.
\newblock In {\em ICSE '10}, 2010.

\end{thebibliography}



\begin{thebibliography}{00}


\ifx \showCODEN    \undefined \def \showCODEN     #1{\unskip}     \fi
\ifx \showDOI      \undefined \def \showDOI       #1{{\tt DOI:}\penalty0{#1}\ }
  \fi
\ifx \showISBNx    \undefined \def \showISBNx     #1{\unskip}     \fi
\ifx \showISBNxiii \undefined \def \showISBNxiii  #1{\unskip}     \fi
\ifx \showISSN     \undefined \def \showISSN      #1{\unskip}     \fi
\ifx \showLCCN     \undefined \def \showLCCN      #1{\unskip}     \fi
\ifx \shownote     \undefined \def \shownote      #1{#1}          \fi
\ifx \showarticletitle \undefined \def \showarticletitle #1{#1}   \fi
\ifx \showURL      \undefined \def \showURL       #1{#1}          \fi
\providecommand\bibfield[2]{#2}
\providecommand\bibinfo[2]{#2}
\providecommand\natexlab[1]{#1}
\providecommand\showeprint[2][]{arXiv:#2}

\end{thebibliography}

\end{document}